\documentclass[pre,twocolumn,amsmath,amssymb,dvipdfmx,superscriptaddress]{revtex4-1}
\usepackage{amsmath,amssymb}
\usepackage{dcolumn}
\usepackage{parskip,comment}
\usepackage{graphicx}
\usepackage[normalem]{ulem}
\usepackage{braket}
\usepackage{here}
\def\0{{\boldsymbol 0}}

\def\l<{\left<} \def\r>{\right>}
\def\wt{\widetilde}

\usepackage[dvips]{color} 

\renewcommand\sout{\bgroup \color{red} \ULdepth=-.5ex \ULset}

\begin{document}
\title{Discreteness-induced transitions in multi-body reaction systems}
\author{Yohei Saito}
\email{yoheis@sat.t.u-tokyo.ac.jp}
\address{Institute of Industrial Science, The University of Tokyo, 4-6-1, Komaba, Meguro-ku, Tokyo 153-8505 Japan}

\author{Yuki Sughiyama}
\address{Institute of Industrial Science, The University of Tokyo, 4-6-1, Komaba, Meguro-ku, Tokyo 153-8505 Japan}
\address{Department of Basic Science, School of Arts and Sciences, The University of Tokyo, 3-8-1, Komaba, Meguro-ku, Tokyo 153-8902, Japan}

\author{Kunihiko Kaneko}
\address{Department of Basic Science, School of Arts and Sciences, The University of Tokyo, 3-8-1, Komaba, Meguro-ku, Tokyo 153-8902, Japan}

\author{Tetsuya J. Kobayashi}
\address{Institute of Industrial Science, The University of Tokyo, 4-6-1, Komaba, Meguro-ku, Tokyo 153-8505 Japan}
\begin{abstract}
A decrease in system size can induce qualitatively different behavior compared to the macroscopic behavior of the corresponding large-size system.  
The mechanisms of this transition, which is known as the {\it small-size transition}, can be attributed 
to either a relative increase in the noise intensity or to the discreteness of the state space due to the small system size. 
The former mechanism has been intensively investigated using several toy and realistic models. However, 
the latter has rarely been analyzed and is sometimes confused with the former, because a toy model that extracts the essence of the discreteness-induced transition mechanism is lacking.
In this work, we propose a $1$- and $3$-body reaction system as a minimal model of the discreteness-induced transition 
and derive the conditions under which this transition occurs in more complex systems.
This work enriches our understanding of the influence of small system size on system behavior. 
\end{abstract}
\maketitle

\section{Introduction}
Detecting qualitative transitions in system's state such as the phases 
is important for characterizing how the system changes its nature depending on the system's parameters. 
For a macroscopic system with an infinite number of particles, 
the $n$-th order phase transition is defined by the existence of a singularity in the $n$-th derivative of the free energy. 
Although a {\it phase} transition with a singularity does not occur in finite or small-size systems, 
the notion of this transition can be extended. 
By considering the typical states of a stochastic finite-size system in conjunction with the peaks of the stationary distribution, 
transitions in the finite-size system can be defined as the emergence or disappearance of peaks in response to changes in the system parameters. 
For example, such a change in the peaks in response to a variation in the noise intensity is known as a noise-induced transition 
\cite{horsthemke2006noise, horsthemke1977phase, lefever1979bistability, suzuki1981phase}. 

Although this phenomenon was reported in the 1970s, 
it has attracted renewed attention in the field of systems biology, because chemical reactions within a cell 
comprising a small number of molecules are typical examples of small-size system behavior. 
Indeed, with a decrease in the system size or in the total number of molecules, 
changes in the distribution peaks appear and a transition due to the small-number effect is exhibited
\cite{togashi2001transitions, togashi2003alteration, ohkubo2008transition, kobayashi2011connection, blumenfeld2012biophysical, biancalani2014noise, PhysRevE.91.022707}. 
This phenomenon is also relevant for evolutionary games in finite populations, for example, 
for the establishment of cooperation and consensus among social agents \cite{nowak2004emergence, nowak2006evolutionary}. 
Despite their prevalence and importance, however, the origins of small-size transitions are not yet fully understood. 
This lack of understanding exists because system smallness has a minimum of two completely different effects on the system: 
it increases the intrinsic noise intensity and renders the state space discrete. 

The original noise-induced transition reported 
in Refs.~\cite{horsthemke2006noise, horsthemke1977phase, lefever1979bistability, suzuki1981phase}, 
along with the majority of other results \cite{ohkubo2008transition, kobayashi2011connection, biancalani2014noise, PhysRevE.91.022707}, 
can be attributed to increased noise intensity and the multiplicative nature of the noise itself. 
This conclusion can be drawn because such transitions can be observed 
even if we employ a continuous approximation of the original dynamics over the discrete state space using the chemical Fokker-Planck equations (CFPE) 
\footnote{In this sense, this phenomenon is appropriately denoted a noise-induced transition.}. 
In this transition, roughly speaking, the system remains in the region for which the noise intensity is smaller than other region in the state space, 
which results in the appearance of a new peak in the stationary distribution.

The dynamic properties and biological implications of noise-induced transitions have already been intensively analyzed 
for both simple toy models \cite{ohkubo2008transition} 
and more complex models \cite{saito2015theoretical}. 
The impact of discreteness, in contrast, has rarely been analyzed. In fact, this factor is not even acknowledged in the majority of works on small-size transitions. 
As an example of a study in which the influence of discreteness is considered, Togashi \textit{et al.} have reported that a dramatic change in dynamics is observed in a small-size autocatalytic reaction system in response to an alternative extinction of molecular species in the autocatalytic loop; this is caused by the discreteness of the system 
\cite{togashi2001transitions, togashi2003alteration}. 
However, although these researchers have identified discreteness as the origin of the observed transition, 
Ohkubo has argued that this transition can instead be attributed to a noise-induced transition, if a simplified version of their model is employed \cite{ohkubo2008transition}. 
Therefore, the ability of system discreteness to induce new transitions remains an uncertain and controversial topic. 

One fundamental problem that hampers our understanding of the role of discreteness 
is the lack of a minimal toy model that extracts the essence of a purely discreteness-induced transition, provided such a transition indeed exists. 
In this paper, we resolve this problem by proposing a $1$- and $3$-body reaction system as a minimal model of the discreteness-induced transition. 
By extending this model, we also derive 
a sufficient condition under which the discreteness-induced transition occurs in more complex systems. 
Finally, we note a possible connection of the system to phenomena other than a reaction system. 
\section{Discreteness-induced transition in $1$- and $3$-body reaction system}
Throughout this paper, we consider spatially homogeneous reaction systems, 
which consist of two species, $A$ and $B$
\footnote{
We assume that the reaction and diffusion time scales are completely separate
and, also, that each reaction occurs instantaneously.}. 
We assume that the total number of particles, $N=n_A+n_B$, is conserved, where $n_A$ and $n_B$ are the number of particles of $A$ and $B$, respectively. 
Thus, the state of the system is determined by the difference in the particle numbers of $A$ and $B$, $z=n_A-n_B$ ($-N\leq z\leq N$), 
and also the system size, $\Omega$, is proportional to $N$, $\Omega\propto N$. 
Henceforth, we call $N$ the system size. 
We also suppose that the number of particles varies by one during each reaction, 
that is, the particle number difference jumps from $z$ to either $z+2$ or $z-2$ for an infinitesimal time step 
\footnote{
In this system, $z=0, \pm2, \pm4, \cdots, \pm N$ for even $N$ 
and $z=\pm1, \pm 3, \cdots, \pm N$ for odd $N$. 
}. 
These assumptions guarantee the detailed balance condition (DBC) at the stationary state (see Appendix A). 
The dynamics of such reaction systems can be described by the master equation (ME), 
 \begin{eqnarray}
  \partial_t P(t,z) 
   &=& \sum_{y=\pm 2} [w(y,z-y)\ P(t,z-y) \nonumber \\
   && -w(y,z)\ P(t,z)] \, , 
  \label{master}
 \end{eqnarray}
where $P(t,z)$ represents the probability that the system is in the state $z$ at time $t$, 
and $w(y,z)$ denotes the transition rate from $z$ to $z+y$. 
As we are not interested in the dependence on the initial conditions, 
we focus on the stationary distribution, $P_{\rm st.}(z)$, 
which can be obtained using the DBC as a recursion relation,  
$P_{\rm st.}(z+y)=[w(y,z)/w(-y,z+y)]\, P_{\rm st.}(z)$, 
and the normalization condition, $\sum_z P_{\rm st.}(z)=1\,$. 

The first model we consider is the following $1$- and $3$-body reaction system: 
 \begin{eqnarray}
  A \xrightarrow{\epsilon} B \, , && B \xrightarrow{\epsilon} A \, ,
  \label{1-reactions} \\
  2A+B \xrightarrow{\lambda_0} 3A \, ,&& A+2B \xrightarrow{\lambda_0} 3B \, , 
  \label{3-reactions-1} \\
  2A+B \xrightarrow{\lambda_0} A+2B \, , && A+2B \xrightarrow{\lambda_0} 2A+B \, , 
  \label{3-reactions-2}   
 \end{eqnarray}
where $\epsilon$ and $\lambda_0$ denote the $1$- and $3$-body reaction rates, respectively 
\footnote{
If the reaction rate of Eq.~(\ref{3-reactions-1}) differs from that of Eq.~(\ref{3-reactions-2}), 
this system may exhibit a second-order phase transition. 
In order to focus on the difference between discreteness- and noise-induced transitions, 
we do not consider systems in which any other transition may occur. However, it is theoretically interesting to evaluate small-size effects in systems with phase transitions. 
}. 
The transition rates are given by 
 \begin{eqnarray}
  w_3(2,z) &=& w_{1,3}(2,z)+w_{3,3}(2,z) \, , 
  \label{tr-A} \\
  w_3(-2,z) &=& w_{1,3}(-2,z)+w_{3,3}(-2,z) \, , 
  \label{tr-B} 
 \end{eqnarray}
where 
 \begin{eqnarray}
  w_{1,3}(\pm 2,z) &=& \frac{\epsilon}{2}\, (N \mp z) \, , 
  \label{tr-A-1} \\
  w_{3,3}(\pm 2,z) &=& \frac{\lambda_0}{4N^2}\, (N+z)(N-z)(N-2) \, .
  \label{tr-A-2} 
 \end{eqnarray}
Here, Eq.~(\ref{tr-A-1}) denotes the transition rates from $z$ to either $z+2$ or $z-2$ 
through $1$-body reactions (Eq.~(\ref{1-reactions})), 
and Eq.~(\ref{tr-A-2}) represents those through $3$-body reactions (Eqs.~(\ref{3-reactions-1}) and (\ref{3-reactions-2})). 
The factor $(N-2)$ in Eq.~(\ref{tr-A-2}) 
means that the $3$-body reactions cannot arise when the total particle number, $N$, is less than $3$ 
\footnote{
At $N=1$, $z$ takes either $1$ or $-1$ and, therefore, the $3$-body transition rates become $0$ owing to the $(N+z)(N-z)$ term. 
}. 
As a result of the DBC, the relation $w(2,z)P_{\rm st.}(z)=w(-2,z+2)P_{\rm st.}(z+2)$ is maintained, 
and the ratio between two neighboring states can be expressed as 
 \begin{eqnarray}
 \frac{P_{\rm st.}(z+2)}{P_{\rm st.}(z)} 
   &=&  \frac{w_3(2,z)}{w_3(-2,z+2)}, \nonumber \\[0.2cm]
   &=& \left[ 1+\frac{f_3(z)}{w_3(-2,z+2)} \right] \, , 
  \label{DBC-3} 
 \end{eqnarray}
 where
  \begin{eqnarray}
  f_3(z) 
   &=& w_3(2,z)-w_3(-2,z+2), \nonumber \\
   &=& \lambda_0 \, (1+z) \, \left[-\frac{\epsilon}{\lambda_0}+\frac{N-2}{N^2}\right] \, .
  \label{eq:f-3}  
 \end{eqnarray}
It is apparent that 
$f_3(z)>0$ and $f_3(z)<0$ correspond to $P_{\rm st.}(z+2)>P_{\rm st.}(z)$ and $P_{\rm st.}(z+2)<P_{\rm st.}(z)\,$, respectively; 
hence, $f_3(z)$ determines the local structure of the stationary distribution. 
Moreover, if the term inside the square bracket in the last line of Eq.~(\ref{eq:f-3}) is negative, 
$P_{\rm st.}(z)$ is convex at $z=0\,$; that is, the stationary distribution is unimodal. 
In contrast, if this term is positive, $P_{\rm st.}(z)$ is concave at $z=0\,$ 
and the stationary distribution becomes bimodal. 
This transition occurs at the $N$ that satisfies $(N-2)/N^2=\epsilon/\lambda_0$, 
and the uniform distribution appears at the stationary state. 
The critical system size is given by 
 \begin{eqnarray}
   N_c^{\pm}(3) &=& \frac{1}{2} \ \left[ \frac{\lambda_0}{\epsilon} \pm \sqrt{\left(\frac{\lambda_0}{\epsilon}\right)^2-\frac{8\lambda_0}{\epsilon}} \ \right] \, ,
  \label{Nc-3body} 
 \end{eqnarray}
where the upper index of the left-hand side corresponds to the sign of the right-hand side. 

\begin{figure}[t]
 \begin{center}
  \includegraphics[width=80mm]{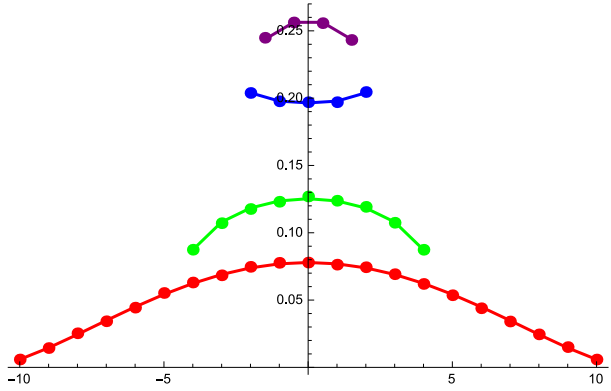}
 \end{center}
 \caption
{Stationary distributions of the $1$- and $3$-body reaction system (Eqs.~(\ref{1-reactions}), (\ref{3-reactions-1}) and (\ref{3-reactions-2})) 
as functions of $z$. 
The curves and points denote the $P_{\rm st.}(z)$, obtained analytically from the DBC 
and numerically from the stochastic simulation by Gillespie's algorithm \cite{gillespie1977exact}, respectively. 
The reaction rates are $\epsilon=12, \lambda=100$, 
and the number of particles is $N=20$ (red), $8$ (green), $4$ (blue), and $3$ (purple). 
The noise-induced transition occurs between $N=8$ and $4$ 
whereas the discreteness-induced transition appears between $N=4$ and $3$. 
}
 \label{fig:3body_distribution}
\end{figure}

\begin{figure}[t]
 \begin{center}
  \includegraphics[width=80mm]{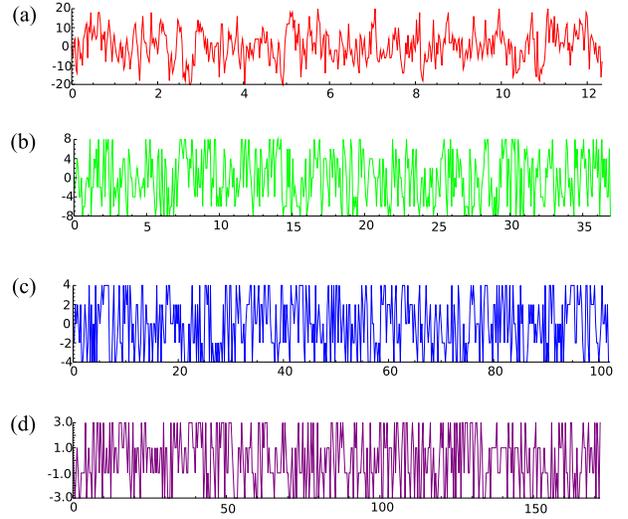}
 \end{center}
 \caption
{Stochastic time evolutions of $1$- and $3$-body reaction system. 
The vertical and horizontal axes denote $z$ and $t$, respectively. 
The correspondence between the colors and parameters, $N,\ \epsilon$, and $\lambda_0$, is the same as in FIG.~\ref{fig:3body_distribution}. 
}
 \label{fig:3body_time}
\end{figure}

We suppose that $N$ decreases from infinity. 
From Eqs.~(\ref{eq:f-3}) and (\ref{Nc-3body}), $P_{\rm st.}(z)$ is unimodal when $N>N_c^+(3)$, 
bimodal when $N_c^-(3)<N<N_c^+(3)$, 
and unimodal for $N<N_c^-(3)\,$. 
FIG.~\ref{fig:3body_distribution} shows the stationary distributions 
obtained analytically from the DBC or numerically from a stochastic simulation. 
We find that two transitions occur in accordance with the decrease in $N$. 
The stochastic trajectories of the conditions shown in FIG.~\ref{fig:3body_time} indicate that the system trapping time is increased as $N$ decreases, 
which causes the transition at $N_c^+(3)$. 
In contrast, the transition at $N_c^-(3)$ cannot be confirmed from the trajectories at $N=4$ (blue) and $3$ (purple), 
unless the stationary distributions, which have small curvatures at their centers, are compared. 

Next, we explain that the transitions at $N_c^+(3)$ and $N_c^-(3)$ can be understood 
as noise- and discreteness-induced transitions, respectively. 
Following Ref.~\cite{ohkubo2008transition}, we divide the reactions, Eqs.~(\ref{tr-A}) and (\ref{tr-B}), 
into symmetric and asymmetric parts, $w_{3,3}(2,z)=w_{3,3}(-2,z)$ and $w_{1,3}(2,z)\neq w_{1,3}(-2,z)\,$, respectively. 
Because of the asymmetricity of the transition rates, $w_{1,3}(2,z)-w_{1,3}(-2,z)=-\epsilon z$, 
one expects that the system at $z$ jumps more frequently toward $z=0$ than $z=\pm N$. 
This transition direction bias can be regarded as a kind of force or drift term in CFPE. 
In contrast, as the symmetric part, $w_{3,3}(2,z)=w_{3,3}(-2,z)$, 
causes the system to jump from $z$ to either $z+2$ or $z-2$ at an equal rate, 
this part works as noise or diffusion term in the CFPE. 
Thus, the dynamics of this system can be visualized via an analogy with a system driven by force and noise. 
From the second line of Eq.~(\ref{eq:f-3}), 
one finds that the first and second terms inside the square bracket represent the drift and noise, respectively. 
Therefore, Eq.~(\ref{eq:f-3}) indicates which effect more strongly determines the structure of $P_{\rm st.}(z)\,$. 
When $N$ is sufficiently large, the noise term can be negligible as a result of the law of large numbers, 
and the stationary behavior is almost fully determined by the drift term. 
In this situation, the system is expected to frequently achieve states in the vicinity of $z=0$, the stable fixed point of the drift term. 
However, the noise term increases as $N$ is decreased, and the noise becomes dominant over the drift. 
That is, even though $z=0$ appears to be stable based on the analysis of the drift term, 
the noise intensity is strongest here; thus, the system is quickly kicked off. 
As a result, one frequently observes the system at states in the vicinity of $z=\pm N$, 
where the noise intensity is small 
or, equivalently, where the system is trapped for a longer period of time 
\footnote{
By decomposing the DBC, 
we can obviously show that the stationary distribution is determined by these two effects. 
However, this interpretation is not the primary topic of this paper. 
In addition, it is not useful to obtain the stationary distribution in practice. 
Thus, this decomposition is not presented in this paper; it will be reported elsewhere. 
}. 
This phenomenon occurs when the noise overcomes the drift term; thus, it is referred to as a noise-induced transition. 

We next evaluate the noise effect when $N$ is further decreased. 
The second term of Eq.~(\ref{eq:f-3}) decreases 
owing to the numerator, ($N-2$). 
Consequently, the drift becomes dominant once more in the case of an extremely small $N$, 
and the unimodal distribution appears. 
The ($N-2$) factor, which is introduced by Eq.~(\ref{tr-A-2}), 
indicates that $3$-body reactions cannot arise when $N<3\,$. 
Hence, it can be concluded that the transition at $N_c^-(3)$ emerges due to the discreteness of $N$. 
In fact, if we approximate ($N-2$) by $N$ and neglect this discreteness effect, the transition at $N_c^-(3)$ disappears. 
Such an approximation is often employed in the Kramers-Moyal expansion 
in order to derive the CFPE, which neglects ${\cal O}(1/N^2)$ terms and fails to describe extremely small-size systems. 
(A detailed calculation is shown in Appendix B.) 

\section{Discreteness-induced transition in $1$- and $M$-body reaction system}
Because of the simplicity of our minimal system, the second critical system size, $N_c^-(3)$, 
is a maximum of 4 when $\frac{\lambda_0}{\epsilon}=8\,$, which is not feasible for any realistic situations. 
In the following, we demonstrate that the critical system size for the discreteness transition can be sufficiently large 
for applications to biological and social systems, by considering general $M$-body reactions rather than those with $M=3$. 
To extract the essence of this behavior analytically, we consider a restricted $M$-body reaction system
\footnote{
For general $M$-body reaction systems, discreteness- and noise-induced transitions may become obscure 
owing to other transitions, {\it e.g.}, the first- and second-phase transitions. 
Therefore, by restricting the $M$-body reaction rates, 
we consider a system in which phase transitions do not arise. 
}: 
 \begin{eqnarray}
  && A \xrightarrow{\epsilon} B \, , \qquad\qquad B \xrightarrow{\epsilon} A \, , 
  \label{1-reactions-2} \\
  && mA+(M-m)B \nonumber \\
  &&\qquad \xrightarrow{\lambda_{M,m}} (m+1)A+(M-m-1)B \, , 
  \label{M-reactions-1} \\
  && mA+(M-m)B \nonumber \\
  &&\qquad \xrightarrow{\lambda_{M,m}} (m-1)A+(M-m+1)B \, , 
  \label{M-reactions-2}   
 \end{eqnarray}
where $m\, (=1,2,\cdots, M-1)$ represents the number of particles that participate in each reaction 
and $\lambda_{M,m}=(M-2)!\times [(m-1)!\,((M-2)-(m-1))! ]^{-1} \, \lambda_0 \equiv {}_{M-2}C_{m-1}\, \lambda_0$ ($\lambda_0>0$) denotes the $M$-body reaction rates. 
The sum of the reaction rates can be written as (see Appendix C) 
 \begin{eqnarray}
  w_M(\pm 2,z) 
   &=& \frac{\epsilon}{2}(N \mp z) \nonumber \\
   && +\frac{\lambda_0}{4 N^{M-1}}\,(N+z)\,(N-z) \prod_{\ell=2}^{M-1} (N-\ell) \, ,  \nonumber \\
   \label{tr-M} 
 \end{eqnarray}
where the products of $(N-\ell)$ indicate that the $M$-body reactions cannot occur when the total particle number is less than $M$. 
Then, we evaluate the increase and decrease of $P_{\rm st.}(z)$ using 
 \begin{eqnarray}
  f_M(z) 
    &=& w_M(2,z) -w_M(-2,z+2), \nonumber \\
    &=& (z+1) \left[ -\epsilon+\frac{\lambda_0}{N^{M-1}} \prod_{\ell=2}^{M-1} (N-\ell) \right], \nonumber \\
    &=& \lambda_0 (z+1) \, \left[-\frac{\epsilon}{\lambda_0}+h(N,M)\right] \, , 
   \label{Pascal-DBC} \\
  h(N,M) &=& \frac{1}{N^{M-1}} \prod_{\ell=2}^{M-1} (N-\ell) \, . 
   \label{h_M}
 \end{eqnarray} 

\begin{figure}[H]
 \begin{center}
  \includegraphics[width=80mm]{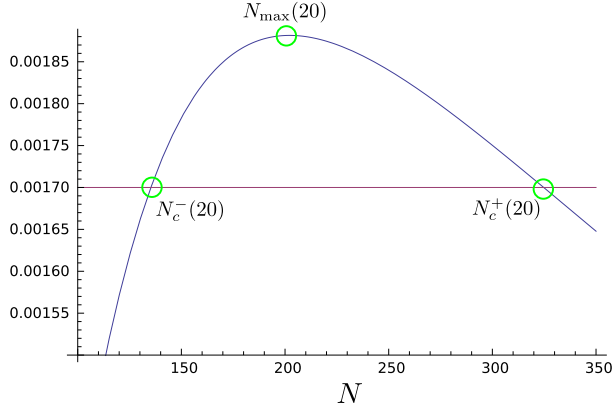}
 \end{center}
 \caption
{First (red) and second (blue) terms inside square brackets of Eq.~(\ref{Pascal-DBC}), $\epsilon/\lambda_0$ and $h(N,M)$, respectively, 
as functions of $N$ for $M=20$, $\lambda_0=10000$, and $\epsilon=17\,$. 
In the region where the blue curve is above (below) the red curve, 
the term inside the bracket is positive (negative), which means that the stationary distribution is concave (convex). 
We find that $h(N,20)$ has a local maximum for $202<N<203$. 
The noise- and discreteness-induced transitions emerge at the intersection points of the two lines, 
which appear after and before this local maximum, respectively. 
}
 \label{fig:1andM}
\end{figure}

As we have explained in the previous model, 
the stationary distribution of this system also changes form from unimodal to bimodal, 
when the sign of the term inside the square bracket in Eq.~(\ref{Pascal-DBC}) changes from negative to positive. 
The critical system size, $N_c(M)$, can be evaluated from $h(N_{c},M)=\epsilon/\lambda_0\,$ 
\footnote{
The product, $\prod_{\ell=2}^{M-1} (N-\ell)$, can be rewritten as $\Gamma(N-1)/\Gamma(N-M+1)$, 
where $\Gamma(x)$ is the gamma function. 
Hence, at $M=2$, $h(N,2)=\lambda_0/\epsilon N$ and the critical system size becomes $N_c(2)=\lambda_0/\epsilon$, 
which has been obtained in previous studies \cite{ohkubo2008transition}. 
Even if we treat this system exactly at $M=2$ using the ME, 
the discreteness-induced transition does not appear. 
}. 
In order to examine both the existence and the number of solutions of this equation, 
it is useful to plot $\epsilon/\lambda_0$ and $h(N,M)$ as functions of $N$, 
and to focus on the local maximum of $h_M(N)\,$ that first appears when $N$ decreases from infinity. 
Let $N_{\rm max}(M)$ be the position of this local maximum. 
In FIG.~\ref{fig:1andM}, we show $h_M(N)$ for $M=20$ as an example, 
and find that this function has a local maximum, $N_{\rm max}$, for $202<N<203\,$. 
We find that, if $\epsilon/\lambda_0<h(N_{\rm max}(M),M)$ is satisfied, two kinds of transition can emerge 
at the $N$ values that satisfy $\epsilon/\lambda_0=h(N_c,M)\,$. 

Next, we suppose $\epsilon/\lambda_0<h(N_{\rm max}(M),M)$ 
and investigate the behavior of $h(N,M)$, which is related to the noise effect, 
when $N$ decreases from infinity. (For instance, see FIG.~\ref{fig:1andM}.) 
First, $h(N,M)$ increases; hence, the noise affects the system. 
As a result, a noise-induced transition arises 
at $N_c^+(M)$, which is the first intersection point of $h(N,M)$ and $\epsilon/\lambda_0$. 
Then, when $N$ is smaller than $N_{\rm max}(M)$, 
$h(N,M)$ is significantly affected by the discreteness of the particle number and the noise effect is again reduced. Consequently, a discreteness-induced transition occurs 
at $N_c^-(M)$, which is the second intersection point of $h(N,M)$ and $\epsilon/\lambda_0$. 
Based on this analysis, we can conclude that 
the critical system size for the discreteness-induced transition, $N_c^-(M)$, can become large 
for most $N_{\rm max}(M)\,$. 
 
\begin{figure}[H]
 \begin{center}
  \includegraphics[width=70mm]{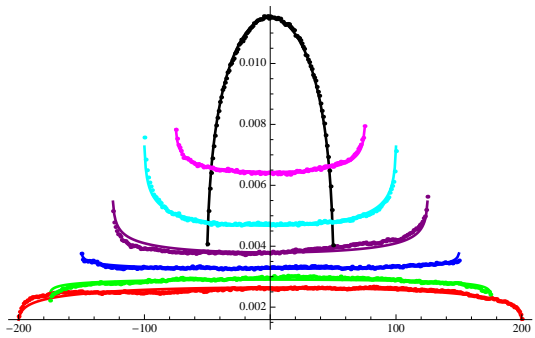}
 \end{center}
 \caption
{Stationary distributions of $1$- and $20$-body reaction system as functions of $z$. 
The curves and points denote the $P_{\rm st.}(z)$ obtained analytically from the DBC 
and numerically from the stochastic simulation \cite{gillespie1977exact}, respectively. 
The reaction rates are $\epsilon=17$, $\lambda_0=10000$, 
and the number of particles is $N=400$ (red), $350$ (green), $300$ (blue), $250$ (purple), 
$200$ (cyan), $150$ (magenta), $100$ (black). 
The noise-induced transition appears between $N=350$ and $300$ 
and the discreteness-induced transition emerges between $N=150$ and $100$.
}
 \label{fig:stationary-5}
\end{figure}

\begin{figure}[t]
 \begin{center}
  \includegraphics[width=90mm]{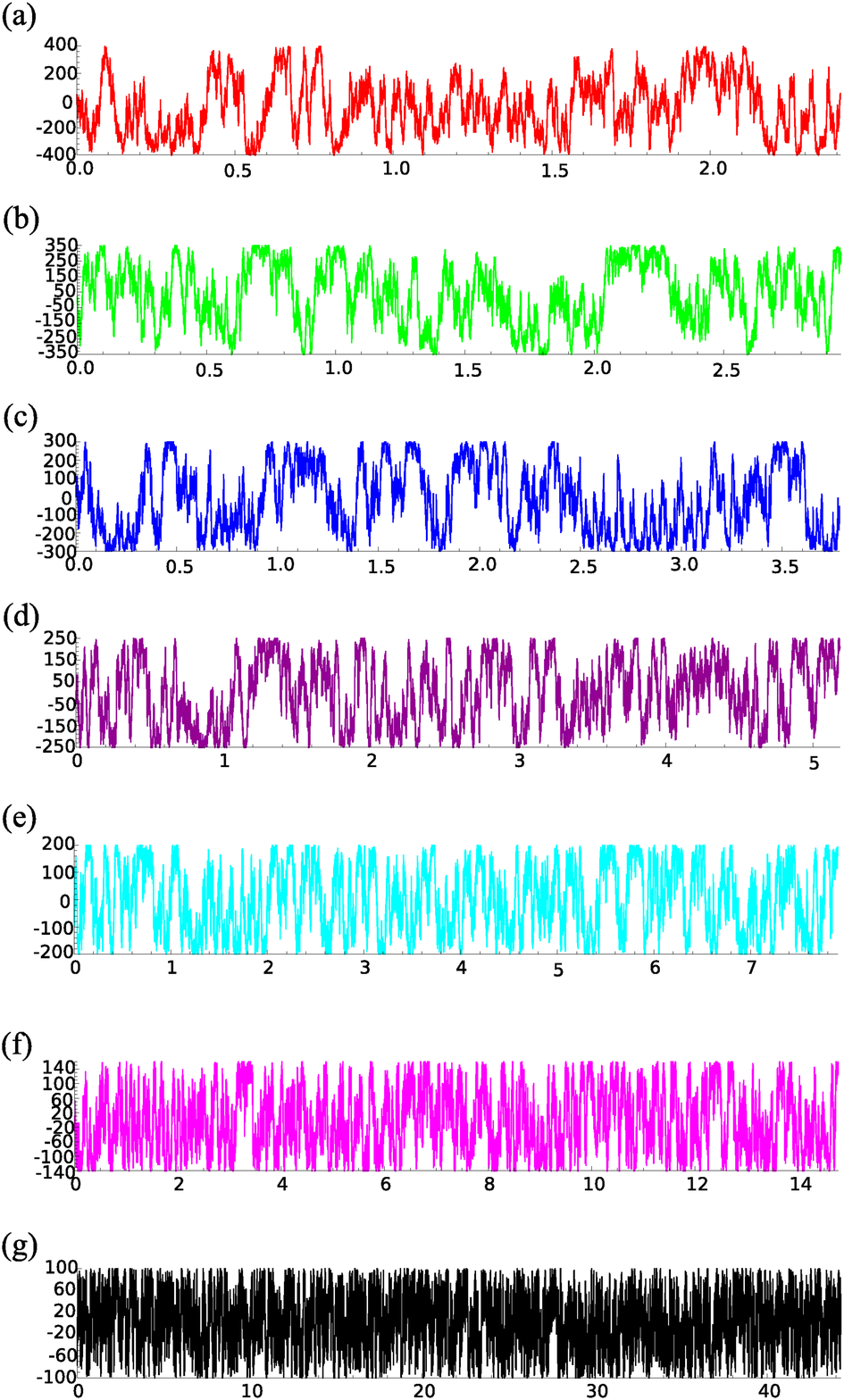}
 \end{center}
 \caption
{Stochastic time evolutions of $1$- and $20$-body reaction system. 
The vertical and horizontal axes denote $z$ and $t$, respectively. 
The correspondence between the colors and parameters, $N,\ \epsilon$, and $\lambda_0$, is the same as in FIG.~\ref{fig:stationary-5}. 
}
 \label{fig:time-5}
\end{figure}

\begin{figure}[H]
 \begin{center}
  \includegraphics[width=70mm]{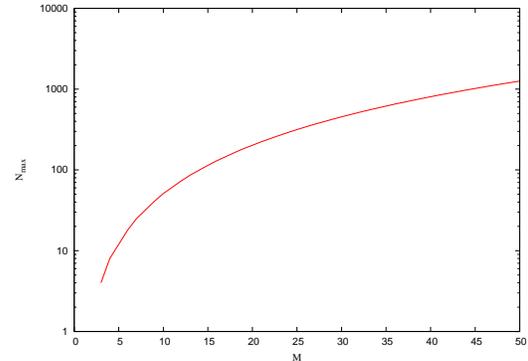}
 \end{center}
 \caption
{Relationship between reaction degree, $M$, 
and particle number $N_{\rm max}$, which becomes the largest extreme point of the second term inside the bracket of Eq.~(\ref{Pascal-DBC}). 
}
 \label{fig:noise-max}
\end{figure}

In FIG.~\ref{fig:stationary-5}, we also confirm that the stationary distribution of this system at $M=20$ 
certainly changes twice across $N_c^+(20)$ and $N_c^-(20)$. 
From the stochastic simulation shown in FIG.~\ref{fig:time-5}, 
we find that, when $N$ is smaller than $N_c^+(20)$, 
the trapping time in the vicinity of the boundaries is increased. 
Furthermore, for $N<N_c^-(20)$, the system appears to frequently 
visits in the vicinity of $z=0$. 
In FIG.~\ref{fig:noise-max}, $N_{\rm max}(M)$ is plotted as a function of $M$ 
and the $N_{\rm max}(M)$ of this system increases quadratically as $M$ becomes large. 
Thus, we conclude that the discreteness-induced transition can be observed for a rather large $N$, 
provided multi-body reactions are permitted. 
Furthermore, if different $M$-body reactions arise in a system, {\it e.g.}, $1$-, $3$-, and $10$-body reactions,  
discreteness-induced transitions may emerge several times as $N$ decreases. 
\section{summary and discussion}
As the size of a system decreases, 
the typical states, which can be associated with the peaks of the stationary distribution, 
may change dramatically, even if no macroscopic transitions emerge. 
The effect of small $N$ is considered to be twofold: 
there is an increase in the intrinsic noise and the state space discreteness is emphasized. 
The transitions investigated in the majority of the previous studies related to this topic
\cite{horsthemke2006noise, horsthemke1977phase, lefever1979bistability, suzuki1981phase, 
ohkubo2008transition, kobayashi2011connection, biancalani2014noise, PhysRevE.91.022707} 
are thought to have been caused by intrinsic noise, 
as the examined systems continued to exhibit transitions even when the state space of the system was assumed to be continuous. 
In contrast, the effect of discreteness has rarely been studied \cite{togashi2001transitions}, 
and the argument that discreteness causes a transition for small $N$ remains controversial. 

In this paper, we have analyzed a $1$- and $3$-body reaction system 
and confirmed the existence of the discreteness-induced transition. 
In addition, we evaluated the critical system size for the discreteness-induced transition 
and showed that, in a restricted $1$- and $M$-body reaction system, 
the critical system size quadratically increases as a function of $M$. 
This result indicates that the discreteness effect may not necessarily be  neglected in finite-size systems. 

Finally, we note that systems with many-body reactions can be interpreted as a consensus model for social agents. 
In such cases, $A$ and $B$ represent two different opinions. 
The $1$-body reactions (Eq.~(\ref{1-reactions-2})) indicate 
individuals spontaneously changing their opinions from $A$ to $B$ or from $B$ to $A$ 
at the rate $\epsilon$, which represents the frequency at which opinions are changed. 
With regard to the $M$-body reactions, the components indicated by Eqs.~(\ref{M-reactions-1}) and (\ref{M-reactions-2}) represent 
an opinion change when $M$-people interact. 
Note that this system is symmetric under the exchange of $A$ and $B$; thus, 
we do not consider the case in which the individuals have an opinion preference. 
As we have shown, the stationary distribution is affected by both the system size, $N$, and the reaction particles, $M$. 
Two peaks appear at $z=\pm N$ in the probability distribution 
for $N_c^-(M)<N<N_c^+(M)\,$. 
These two peaks correspond to the case in which all the individuals have the same opinion, either $A$ or $B$. 
Thus, 
it is expected that a consensus can be achieved without a preferred opinion in this scenario. 
Therefore, when $\epsilon,\, \lambda_0$, and $N$ are given, 
the opinion of a group can be unified by choosing an appropriate convention size, {\it i.e.}, $M$. 

We also comment that our results can be extended to more general systems with no restriction on reaction rates. 
By changing $\epsilon$ and $\lambda_{M,m}$ from those used in this paper, 
we can observe various stationary distributions and a more complex phase diagram for $P_{\rm st.}(z)$. 
In future studies, we will verify noise- and discreteness-induced transitions in spatially inhomogeneous systems without the DBC. 
\begin{acknowledgments}
The authors would like to thank Dr. Nen Saito for his useful comments. 
This research is partially supported by a Grant-in-Aid for Scientific Research on Innovative Areas 
``Molecular Robotics" (No. 24104001-5) from the Ministry of Education, Culture, Sports, Science, and Technology, Japan, the
Platform for Dynamic Approaches to Living Systems funded by MEXT and AMED, Japan, and the JST PRESTO program.
\end{acknowledgments}
\appendix
\section{Detailed balance condition}
We explain here that the systems treated in this paper satisfy the DBC in their stationary states. 
First, consider the ME at a system boundary, $z=N$. 
As the total particle number is conserved in this system, 
the transition rate from $z=N$ to $z=N+2$ is zero, {\it i.e.}, $w(2,N)=0$. 
Thus, at the stationary state, the ME becomes 
 \begin{eqnarray}
  w(-2,N) P_{\rm st.}(N) 
   = w(2,N-2) P_{\rm st.}(N-2) \, . \nonumber \\
  \label{dtb-1}
 \end{eqnarray}
Then, we proceed to the neighboring state, $z=N-2$. 
The ME at this state can be expressed as 
 \begin{eqnarray}	
  &&[w(-2,N-2) +w(2,N-2)] \ P_{\rm st.}(N-2) \nonumber \\
  && =w(2,N-4) P_{\rm st.}(N-4) +w(-2,N) P_{\rm st.}(N) \, . \nonumber \\
  \label{dtb-2}
 \end{eqnarray}
Substituting Eq.~(\ref{dtb-1}) into Eq.~(\ref{dtb-2}), 
we obtain 
 \begin{eqnarray}
  &&w(-2,N-2) P_{\rm st.}(N-2) \nonumber \\
  && = w(2,N-4) P_{\rm st.}(N-4) \, . 
  \label{dtb-3}
 \end{eqnarray}
In the same manner, by considering successive MEs at the neighboring states, 
we can conclude that the DBC 
 \begin{eqnarray}
  w(2,z) P_{\rm st.}(z) &=& w(-2,z+2) P_{\rm st.}(z+2) \, , \nonumber \\
  \label{DBC}  
 \end{eqnarray}
is satisfied in this system. 
\section{Analysis of $3$-body reaction system using chemical Fokker Planck equation}
As it is difficult to analytically solve the ME, even in the stationary state, 
the CFPE is employed to evaluate the properties of the stationary distribution. 
However, the CFPE neglects fluctuations of ${\cal O}(1/N^2)$ and, therefore, 
a qualitatively incorrect result may be derived for small $N$. 
Here, we show that a discreteness-induced transition cannot be found if we employ the CFPE to describe the dynamics of the $1$- and $3$-body reaction system considered in the main text. 
In this Appendix, 
we change the variable from the particle number difference, $z$, 
to the particle concentration difference, $x=z/N$, and assume that $P_{\rm st.}(x)$ is a continuous function. 
This assumption is valid provided $N$ is sufficiently large. 
However, as this approximation neglects the particle-number discreteness, 
it becomes problematic in extremely small-size systems. 
By employing the Kramers-Moyal expansion and neglecting ${\cal O}(1/N^2)$ terms, 
we can derive the CFPE \cite{Gardiner200901} as follows: 
 \begin{eqnarray}
 \begin{split}
  \partial_t P^{\rm CFP} & (t,x) \\
  =& -\partial_x [A(x) P^{\rm CFP}(t,x)] \\
   & +\frac{1}{2N} \ \partial_x^2 \ [B(x) P^{\rm CFP}(t,x)] \, ,
  \label{FP-eq}
 \end{split} 
 \end{eqnarray}
where the drift and diffusion terms, $A(x)$ and $B(x)$, respectively, are given by 
 \begin{eqnarray} 
  A(x) &=& -2\epsilon x \, , \\ 
  B(x) &=& -2\lambda_0 x^2 +2\, [2\epsilon +\lambda_0] \, .
 \end{eqnarray}
Then, the stationary distribution becomes 
 \begin{eqnarray}
  P_{st.}^{\rm CFP}(x) &=& {\cal R}^{-1}\ {\rm e}^{\phi(x)} \, , 
  \label{FP-sol-def} \\
  \phi(x) &=& \left( \frac{N\epsilon}{\lambda_0}-1\right) \, \ln \left( \frac{2\epsilon}{\lambda_0} +1-x^2 \right) \, , \nonumber \\
  \label{FP-sol}
 \end{eqnarray}
where ${\cal R}$ is the normalization factor. 
This system exhibits a noise-induced transition at the critical system size, $N_c=\lambda_0/\epsilon$, 
and this size agrees with the exact result of Eq.~(\ref{Nc-3body}) up to ${\cal O}(1/N)$. 
However, as we neglect ${\cal O}(1/N^2)$ terms, which express the lack of reactants in the $3$-body reactions for small $N$, 
we fail to find discreteness-induced transitions using Eqs.~(\ref{FP-sol-def}) and (\ref{FP-sol}). 
\section{Derivation of Eq.~(\ref{tr-M})}
As the $1$- and $M$-body reaction system treated in the main text is invariant 
under the exchange of $A$ and $B$, 
$w_M(-2,z)$ satisfies $w_M(-2,z)=w_M(2,-z)$ and, therefore, 
it is sufficient to show that the $w_M(2,z)$ given by Eq.~(\ref{M-reactions-1}) becomes 
the second term of Eq.~(\ref{tr-M}). 
For convenience, we rewrite the transition rate through the $M$-body reactions, $w_{M,M}(2,z)$, 
using $n_A$ and $n_B$ in place of $N$ and $z$. We obtain 
 \begin{eqnarray}
  w_{M,M}(2,z) 
   &=& \sum_{m=1}^{M-1} \wt w_{M,m}(2,z), \nonumber \\
   &=& \sum_{m=1}^{M-1} \frac{\lambda_{M,m}}{2^M N^{M-1}} \, \frac{(N+z)!!}{(N+z-2m)!!} \nonumber \\
   && \times \frac{(N-z)!!}{[N-z-2(M-m)]!!}, \nonumber \\
   &=& \sum_{m=1}^{M-1} \frac{\lambda_{M,m}}{N^{M-1}}\, \frac{n_A!}{(n_A+m-M)!} \nonumber \\
   && \times \frac{n_B!}{(n_B-m)!} \, , 
  \label{transition-1} 
 \end{eqnarray}
where $\wt w_{M,m}(2,z)$ denotes the reaction rate of Eq.~(\ref{M-reactions-1}) for each $m$. 
When the reaction rates are given by $\lambda_{M,m}={}_{M-2}C_{m-1}\lambda_0$, 
Eq.~(\ref{transition-1}) becomes 
 \begin{eqnarray}
  &&w_{M,M}(2,z) \nonumber \\
  &&=\frac{\lambda_0}{N^{M-1}} \sum_{m=1}^{M-1} {}_{M-2}C_{m-1} \, \frac{n_A!}{(n_A+m-M)!} \, \frac{n_B!}{(n_B-m)!}, \nonumber \\
  &&\equiv \frac{\lambda_0}{N^{M-1}} \, g_M(n_A,n_B) \, .
 \end{eqnarray}
This can be simplified as follows:   
 \begin{eqnarray} 
  &&g_M(n_A,n_B) \nonumber \\
  &&=n_A(n_A-1)\cdots (n_A-M+2) n_B \nonumber \\
  &&+ {}_{M-2}C_{1} \, n_A(n_A-1)\cdots (n_A-M+3) \nonumber \\
  && \qquad \times n_B(n_B-1) \nonumber \\
  &&+ {}_{M-2}C_{2} \, n_A(n_A-1)\cdots (n_A-M+4) \nonumber \\
  && \qquad \times n_B(n_B-1)(n_B-2) \nonumber \\
  &&+\cdots \nonumber \\
  &&+ {}_{M-2}C_{m-1} \, n_A(n_A-1)\cdots (n_A-M+m+1) \nonumber \\
  && \qquad\qquad \times n_B(n_B-1) \cdots (n_B-m+1) \nonumber \\
  &&+\cdots \nonumber \\
  &&+ {}_{M-2}C_{M-2} \, n_A\, n_B(n_B-1)\cdots (n_B-M+2), \nonumber \\[0.3cm]
  &&=n_A\cdots (n_A-M+3)\, n_B \nonumber \\
  && \qquad \times  [(n_A-M+2)+(n_B-1)] \nonumber \\
  &&+[{}_{M-2}C_{1}-1]\,n_A\cdots (n_A-M+4)\, n_B(n_B-1) \nonumber \\
  && \qquad\qquad \times [(n_A-M+3)+(n_B-2)] \nonumber \\
  &&+\cdots \nonumber \\
  &&+[{}_{M-2}C_{m-1}-{}_{M-2}C_{m-2}+{}_{M-2}C_{m-3}+\cdots] \nonumber \\
  &&\quad \times n_A \cdots (n_A-M+m+2) \, n_B\cdots (n_B-m+1) \nonumber \\
  &&\qquad\qquad \times [(n_A-M+m+1)+(n_B-m)] \nonumber \\
  &&+\cdots, \nonumber \\[0.3cm]
  &&=\sum_{m=1}^{M-1} \left( \sum_{\ell =0}^{m-1} (-1)^{m-1-\ell} \, {}_{M-2}C_{\ell} \right) \nonumber \\
  &&\quad \times \frac{n_A!}{(n_A+m-M+1)!} \, \frac{n_B!}{(n_B-m)!} \times [N-M+1] \, , \nonumber \\
  \label{eq:M-body-poly}
 \end{eqnarray}
where we use $n_A+n_B=N$ in the last line of the above equation. 
The term within parentheses in Eq.~(\ref{eq:M-body-poly}) can be rewritten into a simple form. 
Taking the alternating sum of ${}_{M-3}C_{\ell}+{}_{M-3}C_{\ell-1}={}_{M-2}C_{\ell}$ 
from $\ell=m-1$ to $0$ (note that ${}_{M-3}C_{-1}=0$), 
we obtain 
 \begin{eqnarray}
  {}_{M-3}C_{m-1}
   &=& {}_{M-2}C_{m-1}-{}_{M-2}C_{m-2}+{}_{M-2}C_{m-3} \nonumber \\
   &&+\cdots +(-1)^{m-1} {}_{M-2}C_{0}, \nonumber \\
   &=& \sum_{\ell=0}^{m-1} (-1)^{m-1-\ell} \, {}_{M-2}C_{\ell} \, . 
  \label{eq:trans-conb} 
 \end{eqnarray}
Therefore, Eq.~(\ref{eq:M-body-poly}) becomes 
 \begin{eqnarray}
  &&g_M(n_A,n_B) \nonumber \\
  &&=[N-M+1] \, \sum_{m=1}^{M-2} {}_{M-3}C_{m-1} \nonumber \\
  &&\quad \times \frac{n_A!}{(n_A+m-M+1)!} \, \frac{n_B!}{(n_B-m)!}, \nonumber \\
  &=& [N-M+1] \, g_{M-1}(n_A,n_B) \, ,
  \label{eq:M-body-poly-2}
 \end{eqnarray}
where we change the upper boundary of the summation from $M-1$ to $M-2$ using ${}_{M-3}C_{M-2}=0$. 
By applying the same transformation repeatedly, we find 
 \begin{eqnarray}
  g_M(n_A,n_B) 
   &=& [N-M+1] \, [N-M+2] \times \cdots \nonumber \\
   && \times [N-2] \, g_2(n_A,n_B), \nonumber \\
   &=& \prod_{\ell=2}^{M-1} [N-\ell] \, n_A n_B \, .
 \end{eqnarray}
Finally, we obtain the second term of Eq.~(\ref{tr-M}), 
 \begin{eqnarray}
  w_{M,M}(2,z) 
  &=& \frac{\lambda_0}{N^{M-1}} \, \prod_{\ell=2}^{M-1} [N-\ell] \, n_A n_B, \nonumber \\
  &=& \frac{\lambda_0}{4N^{M-1}} \, \prod_{\ell=2}^{M-1} [N-\ell] \nonumber \\
  && \times (N+z)(N-z) \, .
  \label{eq:M-body-poly-3}  
 \end{eqnarray}
The other part of the $M$-body reaction rate, $w_{M,M}(-2,z)$, can be derived from $w_{M,M}(-2,z)=w_{M,M}(2,-z)$, 
as we have noted at the beginning of this Appendix. 
\bibliography{biblio}
\end{document}